\documentclass[sigconf]{acmart}

\AtBeginDocument{%
  }
\usepackage{multirow} 
\usepackage{adjustbox}
\usepackage{placeins}
\usepackage{algorithm} 
\usepackage{algpseudocode}

\setcopyright{acmlicensed}
\copyrightyear{2018}
\acmYear{2018}
\acmDOI{XXXXXXX.XXXXXXX}
\acmConference[Conference acronym 'XX]{Make sure to enter the correct
  conference title from your rights confirmation email}{June 03--05,
  2018}{Woodstock, NY}
\acmISBN{978-1-4503-XXXX-X/2018/06}

\begin{document}

\title{Bidirectional Knowledge Distillation for Enhancing Sequential Recommendation with Large Language Models}

\author{Jiongran Wu}
\authornote{Equal contribution.}
\author{Jiahao Liu}
\authornotemark[1]
\affiliation{
  \institution{Fudan University}
  \city{Shanghai}
  \country{China}
}
\email{jrwu23@m.fudan.edu.cn}
\email{jiahaoliu23@m.fudan.edu.cn}


\author{Dongsheng Li}
\affiliation{
  \institution{Microsoft Research Asia}
  \city{Shanghai}
  \country{China}
}
\email{dongsli@microsoft.com}

\author{Guangping Zhang}
\affiliation{
  \institution{Fudan University}
  \city{Shanghai}
  \country{China}
}
\email{gpzhang20@fudan.edu.cn}

\author{Mingzhe Han}
\affiliation{
  \institution{Fudan University}
  \city{Shanghai}
  \country{China}
}
\email{mzhan22@m.fudan.edu.cn}

\author{Hansu Gu}
\affiliation{
  \institution{Independent}
  \city{Seattle}
  \country{United States}
}
\email{hansug@acm.org}

\author{Peng Zhang}
\affiliation{
  \institution{Fudan University}
  \city{Shanghai}
  \country{China}
}
\email{zhangpeng_@fudan.edu.cn}

\author{Li Shang}
\affiliation{
  \institution{Fudan University}
  \city{Shanghai}
  \country{China}
}
\email{lishang@fudan.edu.cn}

\author{Tun Lu}
\affiliation{
  \institution{Fudan University}
  \city{Shanghai}
  \country{China}
}
\email{lutun@fudan.edu.cn}

\author{Ning Gu}
\affiliation{
  \institution{Fudan University}
  \city{Shanghai}
  \country{China}
}
\email{ninggu@fudan.edu.cn}

\newcommand{\myString}{LLMD4Rec}
\newcommand{\myStringC}{{\myString$_C$}}
\newcommand{\myStringL}{{\myString$_L$}}

\renewcommand{\shortauthors}{Jiongran Wu et al.}

\begin{abstract}
Large language models (LLMs) have demonstrated exceptional performance in understanding and generating semantic patterns, making them promising candidates for sequential recommendation tasks. However, when combined with conventional recommendation models (CRMs), LLMs often face challenges related to high inference costs and static knowledge transfer methods. In this paper, we propose a novel mutual distillation framework, LLMD4Rec, that fosters dynamic and bidirectional knowledge exchange between LLM-centric and CRM-based recommendation systems. Unlike traditional unidirectional distillation methods, LLMD4Rec enables iterative optimization by alternately refining both models—enhancing the semantic understanding of CRMs and enriching LLMs with collaborative signals from user-item interactions. By leveraging sample-wise adaptive weighting and aligning output distributions, our approach eliminates the need for additional parameters while ensuring effective knowledge transfer. Extensive experiments on real-world datasets demonstrate that LLMD4Rec significantly improves recommendation accuracy across multiple benchmarks without increasing inference costs. This method provides a scalable and efficient solution for combining the strengths of both LLMs and CRMs in sequential recommendation systems.
\end{abstract}



\maketitle

\section{Introduction}
\begin{figure*}[htbp]      
\centering      
\includegraphics[width=0.95\textwidth]{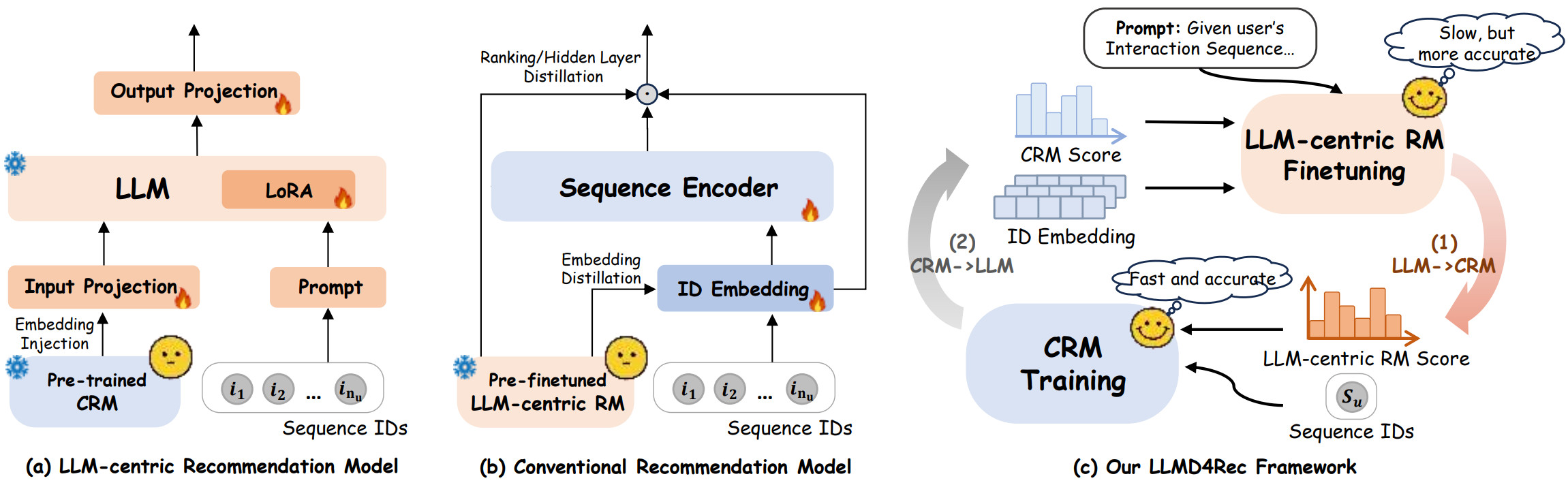}     
\caption{ 
The illustration of the structural comparison between and LLM-centric RM and CRM distillation, along with our \myString's process of mutual distillation training using both components.
In (a), the pre-trained CRM supplies item ID embeddings, while the LLM-centric RM is optimized by fine-tuning its LoRA modules and projection layers.
In (b), the pre-finetuned LLM-centric RM provides representations from its embedding or top-K ranking result to guide the training and optimization of the CRM.
In (c), the top and bottom information from the CRM is utilized to guide the fine-tuning of the LLM, while the more accurate recommendation scores generated by the LLM-centric RM further contribute to optimizing the recommendation performance of the CRM.
} \label{fig:process}  \end{figure*}

Large language models (LLMs) have demonstrated significant success across various domains, leveraging their exceptional abilities in understanding, reasoning, and generation~\cite{sun2024adaptive,zhang2024agentcf,lu2024aligning,liu2025enhancing1,liu2025enhancing2,liu2025filtering,gu2025llm}. Consequently, the application of LLMs in sequential recommendation (SR), which involves predicting the next item in a user's interaction sequence based on historical data, has attracted growing attention~\cite{liao2024llara,li2023e4srec,liu2023triple,liu2023autoseqrec,liu2022parameter}. Unlike conventional recommendation models (CRMs), which model sparse interaction features directly~\cite{kang2018self,hidasi2015session,liu2023recommendation,liu2023personalized,liu2025mitigating,han2025fedcia}, LLMs can extract deep semantic associations from contextual information, opening up new possibilities for user modeling and recommendation generation~\cite{liao2024llara,li2023e4srec}.

Based on the direction of knowledge transfer, LLMs contribute to SR in two ways. \textbf{(1) LLM → CRM.} Here, LLMs enhance CRM by serving as an encoder for user or item representations, or as an external knowledge source for semantic enhancement~\cite{hu2024enhancing, harte2023leveraging, liu2024llm, xi2024towards}. A typical method in this approach is knowledge distillation (KD), a model-agnostic technique that transfers knowledge from a large pre-trained teacher model to a smaller student model~\cite{gou2021knowledge}. This accelerates the training process and improves model performance. After training, only the student model is used for inference, significantly improving efficiency. As shown in Figure~\ref{fig:process} (a), existing methods first train a large teacher model based on LLM, then transfer knowledge from the embedding layer~\cite{cui2024distillation}, hidden layers~\cite{liu2024large}, or final Top-k results~\cite{du2025active} to a smaller student model based on CRM. 
Post-KD training, the student model's performance moves closer to that of the teacher model, while inference latency is greatly reduced due to its smaller size. 
\textbf{(2) CRM → LLM}: In this case, LLMs directly perform recommendation tasks, enabling end-to-end recommendation via prompt learning or parameter-efficient fine-tuning methods such as LoRA~\cite{li2023prompt,liao2024llara,li2023e4srec}. Modal transformation is a common approach. As shown in Figure~\ref{fig:process} (b), these methods~\cite{liao2024llara,li2023e4srec} first train a CRM that models the collaborative information between items based on user interaction patterns. Then, they use a input projection layer to transfer embeddings learned from the CRM into the LLM space, thereby transferring collaborative knowledge to the LLM. Finally, the next token prediction layer in the LLM is replaced with a output projection layer to complete the sequential recommendation task. These methods have demonstrated significant advantages on several public benchmark datasets for recommendation. 

Using KD to transfer knowledge from LLM to CRM (LLM → CRM) can improve the accuracy of CRM without increasing inference costs, as inference does not rely on LLMs. Conversely, using LLM as the recommendation models (LLM-centric RMs) and transferring knowledge from CRM to LLM (CRM → LLM) can result in a greater accuracy improvement over CRM, but at the cost of increased inference burden. In this paper, we propose a mutual distillation framework between LLM-centric RM and CRM named \textbf{\myString} (LLM mutual distillation for rec), which combines the benefits of both LLM → CRM and CRM → LLM paradigms, enabling the model to match or even exceed the accuracy of LLM-centric RM without increasing inference costs. The proposed framework is built upon the identification and resolution of two key shortcomings in existing methods, specifically:

\textbf{Firstly, the two existing knowledge transfer paradigms are unidirectional and static}, with knowledge transferred either from LLM to CRM (LLM → CRM) or from CRM to LLM (CRM → LLM). 
In this process, the source of knowledge remains static, receiving no updates. 
For instance, in CRM → LLM, embeddings with collaborative semantics are frozen during LLM training, with only minor adjustments made via projection layers. Intuitively, a better CRM transfers higher-quality knowledge to the LLM, improving it, and vice versa. As shown in Figure~\ref{fig:process} (c), in contrast to existing unidirectional distillation paradigms that designate LLM-centric RM and CRM as the student or teacher, we abstracted two knowledge distillation processes: \textbf{(1) Downward Enhancement for LLM (CRM → LLM)}, and \textbf{(2) Upward Semantics Distillation for CRM (LLM → CRM)}. By iteratively optimizing the LLM and CRM through these two steps, an optimized LLM-centric RM guides the CRM, which then feeds back to enhance the LLM-centric RM further. This process enables dynamic collaborative optimization. Knowledge is transferred in a \textbf{CRM → LLM → CRM → ... → CRM} cycle, enabling bidirectional learning and collaborative optimization between the two models. 

\textbf{Secondly, the two existing knowledge transfer paradigms fail to transfer sufficient knowledge and do not fully capture the global knowledge of the source model.} In deep learning models, layers closer to the output capture more comprehensive semantic information~\cite{liu2024large}.
In the CRM → LLM direction, current methods mostly transform the embedding layer, while a few align user representations, yet both types of approaches neglect other knowledge-bearing components in CRM, such as the prediction layers.
For LLM → CRM, existing distillation approaches—applied to the \textbf{embedding layer}, \textbf{hidden layers}, and \textbf{Top-K results}—also fall short. Specifically: (1) Hidden layer distillation often focuses on those near the output but requires feature transformations due to space mismatches, potentially leading to "cheating"~\cite{zhu2025preference}; (2) In CRM (e.g., SASRec), the prediction layer often shares parameters with item embeddings, unlike in LLM-centric RMs, so hidden-layer distillation fails to transfer final prediction knowledge. (3) Distilling \textbf{Top-K results} only transfers partial ranking information and loses knowledge from raw score distributions.
Given that the output layers of both LLM and CRM reside in the same space and encapsulate knowledge from upstream layers, we propose \textbf{using the prediction layer distribution as the distillation target in both directions}, eliminating the need for additional transformations. We also introduce a \textbf{sample-wise adaptive weighting mechanism} to better leverage the teacher’s knowledge in both transfer directions.

\myString{}aligns the recommendation scoring space of CRMs with that of LLMs, enhancing the smaller model's ability to capture high-order semantic relationships and generate more refined ID embedding representations. Unlike previous methods limited to injecting fixed ID embeddings, this framework allows the LLM and CRMs to mutually leverage each other's low-level embeddings and high-level outputs, enhancing information richness and recommendation accuracy. By introducing a bidirectional knowledge transfer mechanism, the framework establishes a dynamic and complementary learning relationship between CRMs and LLMs, facilitating mutual enhancement and joint optimization during training. The experimental results demonstrate that our method significantly improves the recommendation performance of both CRM and LLM-centric RMs across multiple sequential recommendation datasets, while also accelerating the convergence of LLM-centric RMs, exhibiting strong practicality and scalability.

The main contributions of this paper are as follows:
\begin{itemize}
 \item {} We propose a mutual distillation framework enabling dynamic interaction and cooperative learning between LLM-centric models and conventional models, requiring no additional trainable parameters.

\item {} We introduce the distillation strategy that transfers knowledge through output distribution matching, and further enhance the learning process with a sample-wise adaptive weighting mechanism.

\item {} We conducted extensive experiments on multiple real-world recommendation datasets to verify the significant improvements of the proposed method in terms of both accuracy and efficiency.
\end{itemize}

\section{Related Work}
\subsection{Sequential Recommendation}
Early sequential recommendation methods typically employed Markov Chains (MC) to model sequential patterns~\cite{shani2005mdp,he2016fusing}. With the advent of deep learning, Recurrent Neural Networks (RNNs) and their variants, such as Long Short-Term Memory (LSTM) and Gated Recurrent Units (GRU) were adopted to capture user behavior sequences~\cite{smirnova2017contextual,xu2019recurrent,xu2021long,hidasi2015session}. 
These approaches leverage recurrent structures to model temporal dependencies in user interactions, enabling step-by-step updates of user states for next-item prediction. 
As the field evolved, attention mechanisms and Transformer-based architectures were introduced into sequential recommendation tasks~\cite{kang2018self,de2021transformers4rec,li2020time}. 
For instance, Wang et al.~\cite{kang2018self} proposed SASRec, a self-attention-based model that adaptively selects relevant past interactions to predict the next item, combining the strengths of Markov Chains and RNNs. 
Moreira et al.~\cite{de2021transformers4rec} proposed Transformers4Rec, which incorporates the Transformer architecture and training strategies into sequential recommendation, further advancing the state of the art.

With the rising adoption of LLMs within sequential recommendation systems, some approaches leverage LLMs to enhance CRMs by incorporating them as auxiliary knowledge sources~\cite{xi2024towards} or as user/item encoders~\cite{hu2024enhancing, harte2023leveraging, liu2024llm} to enrich representation learning. 
Another line of approaches directly employs LLMs as recommenders, aiming to improve user understanding by learning interactions and context. 
For example, LLM-TRSR~\cite{zheng2024harnessing} uses hierarchical text summarization to compress user history, enhancing LLM performance in text-rich sequential recommendation. 
Methods like E4SRec~\cite{li2023e4srec} and Llara~\cite{liao2024llara} integrate LLMs with conventional recommender by feeding ID sequences into frozen models with minimal trainable parameters. While these approaches are effective and efficient, they rely mainly on item IDs and underutilize rich structural information. Moreover, existing LLM-centric methods still face deployment challenges such as slow inference and high computational cost~\cite{xu2024slmrec}.

\subsection{Knowledge Distillation}
Knowledge distillation (KD) is an effective model compression technique, in which a smaller student model is trained under the guidance of the output distribution generated by a larger teacher model~\cite{hinton2015distilling}. 
A common approach involves training the student model using the teacher’s soft labels, which offer richer inter-class relational information compared to hard labels. 
Building on this, some studies have introduced intermediate feature alignment to enrich the student model’s representations~\cite{romero2014fitnets,yim2017gift}. 
Others have explored enhancements to the KD loss function, including adversarial training~\cite{xu2017training} and the use of alternative losses like hinge loss~\cite{heo2019knowledge} to better preserve the teacher model’s decision boundaries. 
To overcome the limitations of the fixed, one-way teacher–student structure and enhance collaborative learning between models, Zhao et al.~\cite{zhao2021mutual} proposed a trainable mutual distillation framework that achieves superior end-to-end performance compared to conventional KD approaches. It has found successful applications in domains such as Trajectory-User Linking (TUL) and Spoken Language Understanding (SLU) ~\cite{chen2022mutual,cheng2023ml}.

Owing to the demonstrated effectiveness of LLM-centric RMs, several studies have adopted these high-performance albeit high-latency models as the target for knowledge distillation~\cite{zhang2024delrec,liu2025cora,liu2024large,cui2024distillation}. 
One strand of research emphasizes the incorporation of CRM information into LLM-centric RM, positioning CRM as teacher and LLM as student. 
These approaches aim to optimize the performance of LLM-centric RMs through mediums such as prompts~\cite{zhang2024delrec}, LoRA weight matrices~\cite{liu2025cora} and user representation~\cite{kim2025lost}.
Conversely, other research focus on leveraging more powerful LLM-centric RM to enhance the performance of more efficient CRM, treating LLM as teacher and CRM as student.
For example, Liu et al.~\cite{liu2024large} employed feature-level distillation, directly projecting the hidden layers of LLMs into the output space of conventional models, thereby facilitating knowledge transfer across feature spaces. Cui et al.~\cite{cui2024distillation}, on the other hand, utilized embedding-level distillation, assessing sample importance via position-aware weights, confidence-aware weights, and consistency-aware weights.
To improve efficiency, ALKDRec~\cite{du2025active} introduces selective instance distillation, prioritizing informative samples and filtering out incorrect or redundant LLM predictions to maximize distillation gains.
These studies validate that knowledge can be transferred bidirectionally between large models and conventional models. However, the majority of these methodologies continue to employ static, unidirectional teacher-student distillation frameworks, which may not fully exploit the synergistic potential between different models.

\section{Preliminaries}

\subsection{Task Definition}
In sequential recommendation tasks, let \( \mathcal{U} = \{u_1, u_2, \ldots, u_{\mathcal{|U|}}\} \) denote the set of users and \( \mathcal{I} = \{i_1, i_2, \ldots, i_{\mathcal{|I|}}\} \) denote the set of items. 
For a user \( u \in \mathcal{U} \), the interaction sequence is represented as \( S_u = (i^{(u)}_1,i^{(u)}_2, \ldots, i^{(u)}_t, \ldots, i^{(u)}_{n_u}) \), where \( i^{(u)}_t \in \mathcal{I} \) indicates the item interacted with by user \( u \) at time step \( t \), and \( n_u \) is the length of user \( u \)'s interaction sequence. 
Given the interaction history \( S_u \), the goal of sequential recommendation is to predict the next item the user will interact with at time step \( n_u + 1 \), i.e., to model the probability \( p(i^{(u)}_{n_u+1} \mid S_u) \).

\subsection{Conventional Sequential Recommendation}

Conventional sequential recommendation models, such as SASRec~\cite{kang2018self} and GRU4Rec~\cite{hidasi2015session}, generally follow a unified architecture that consists of three main components.

\textbf{\textbf{1. Embedding Layer.}} Each historical item is mapped into a dense vector using an embedding matrix. The item embedding at time step \( t \) is given by:
\[
\mathbf{e}^{(u)}_t = \mathbf{E}\left[i^{(u)}_t\right] \in \mathbb{R}^d, \quad \forall t = 1, \dots, n_u\text{.}
\]
The entire sequence is represented as
\[
\mathbf{X}_u = [\mathbf{e}^{(u)}_1, \mathbf{e}^{(u)}_2, \dots, \mathbf{e}^{(u)}_{n_u}] \in \mathbb{R}^{n_u \times d}.
\]

\textbf{\textbf{2.Sequence Encoder.}} It captures user dynamics and sequential patterns. It encodes the item embedding sequence into a fixed-length user representation:
\[
\mathbf{h}_u = \text{SeqEncoder}(\mathbf{X}_u) \in \mathbb{R}^d.
\]

\textbf{\textbf{3. Prediction layer.}} Each candidate item \( i \in \mathcal{I} \) is scored based on its similarity to the user's representation. A common scoring function is the dot product:
\[
\hat{y}_c = \mathbf{h}_u^\top \cdot \mathbf{E}[i].
\]

This scoring mechanism can be interpreted as defining a probability distribution over all items conditioned on the user's interaction sequence:
\[
p_c(i^{(u)}_{n_u+1} \mid \mathbf{S}_u) = \text{Softmax}(\hat{y}_c)
\]

To train the model, a loss function is defined to encourage the predicted score of the positive item to be higher than those of negative items. A general training objective can be written as:
\[
\mathcal{L}_{\text{rec}}^{\text{CRM}} = \sum_{u \in U} \sum_{t=1}^{n_u - 1} \ell(\hat{y}_{i^{(u)}_{t+1}}, \hat{y}_{v^{-}}),
\]
where \( \hat{y}_{i^{(u)}_{t+1}} \) is the score of the ground-truth next item, \( v^{-} \) is a sampled negative item, and \( \ell(\cdot) \) denotes a pairwise or pointwise loss function. These objectives are optimized over the entire training set using stochastic gradient descent or its variants.




\subsection{LLM-centric Sequential Recommendation}  

With the rapid development of LLMs, recent works have explored their integration into sequential recommendation tasks. In this work, we adopt \textbf{E4SRec} as a representative LLM-centric sequential RM due to its effectiveness, simplicity, and generality. E4SRec uses a frozen LLM as a feature extractor and learns user representations for prediction, making it a typical embedding-based LLM recommender. It has shown strong performance across multiple benchmarks compared to conventional methods and has been widely extended in follow-up studies~\cite{xu2024slmrec,liu2024large}. Its modular design also facilitates adaptation, analysis, and distillation.

Specifically, E4SRec comprises item ID embeddings $\mathbf{E}$, an input projection layer $W_{\text{in}}$, LoRA adapters, and an output projection layer $W_{\text{out}}$. These components are decoupled from the LLM backbone, enabling better scalability and adaptability. The core modules are described as follows:

\textbf{1. Input Layer.}  
The input layer transforms the CRM's item ID embeddings $\mathbf{E} \in \mathbb{R}^{|\mathcal{I}| \times d}$ into a format compatible with the LLM, where $|\mathcal{I}|$ is the number of items and $d$ is the embedding dimension. Specifically, these embeddings are then linearly projected via $W_{\text{in}} \in \mathbb{R}^{D \times d}$ to match the token dimension $D$ of the LLM:
\begin{equation}
\mathbf{F}_{\text{proj}} =
W_{\text{in}} \cdot \mathbf{E} \in \mathbb{R}^{|\mathcal{I}| \times D}
\label{eq:injection}
\end{equation}

\textbf{2. LLM Layer.}  
In this layer, E4SRec converts the user interaction sequence $\mathbf{S}_u$ into a natural language prompt $\mathbf{P}_u$ and feeds it to a pre-trained LLM (e.g., Qwen, Llama). The LLM encoder then processes the input to generate a contextualized representation of the user's interests. Following standard practice, the final hidden state corresponding to the last item token is used as the user representation:
\[    
    \mathbf{g}_u = \text{LLMEncoder}(\mathbf{P}_u, \mathbf{F}).
\]

\textbf{3. Prediction Layer.} 
The user representation $\mathbf{g}_u$ is then projected to a score vector over all candidate items using a learnable output projection matrix $W_{\text{out}} \in \mathbb{R}^{D \times |I|}$:
\[
    \hat{y}_{l} = \mathbf{g}_u^\top W_{\text{out}},
\]
where $\hat{y}_{l} \in \mathbb{R}^{|I|}$ contains the unnormalized preference scores. The final probability distribution over items is computed using the softmax function:
\[
    p_l(i^{(u)}_{n_u+1} \mid \mathbf{S}_u) = \text{Softmax}(\hat{y}_{l}).
\]

E4SRec is trained using the cross-entropy loss, where each user sequence is split into multiple prefix-target pairs for supervision. The loss encourages the model to assign higher scores to the ground-truth item among all candidates. Formally, the training objective is defined as: 
\[ 
\mathcal{L}_{\text{rec}}^{\text{LLM}} = - \sum_{u \in U} \sum_{t=1}^{n_u - 1} \log p_l(i^{(u)}_{t+1} \mid \mathbf{S}_u^{(t)}), 
\]
where \( \mathbf{S}_u^{(t)} = [i^{(u)}_1, \dots, i^{(u)}_t] \) is the interaction prefix of user \( u \) up to step \( t \).

\begin{figure*}[htbp]     
\centering     
\includegraphics[width=0.9\textwidth]{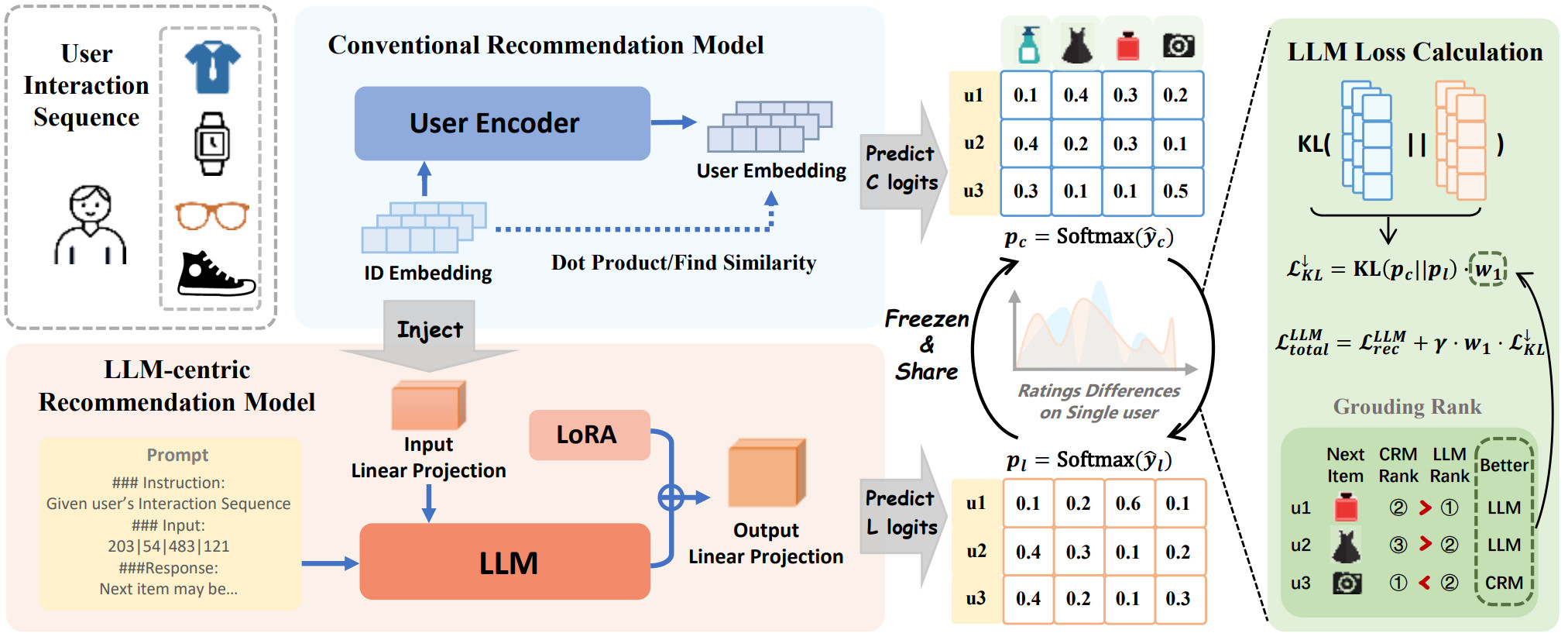}    
\caption{Overview of the proposed \myString{} framework, depicting the directional flow of information and the iterative update mechanism of the LLM-centric RM. The update mechanism for the CRM is similar.
}     
\label{fig:wide-image} \end{figure*}

\section{\myString}
In this section, we present the implementation details of the \myString.

\subsection{Framework Overview}
Our mutual distillation framework establishes a bidirectional learning paradigm between CRMs and LLM-centric RMs. Distinguished from previous studies that rely on unidirectional knowledge transfer or fixed embedding injection, our method alternates training between the two models, dynamically exchanging distilled recommendation signals to enhance performance.



As illustrated in Figure~\ref{fig:wide-image}, the framework proceeds through multiple stages. Initially, the CRM is trained independently using standard sequential recommendation objectives to learn robust item ID embeddings and user representations. 
These embeddings are then injected into the LLM-centric RM, and the probability distributions over items from the CRM serve as initial distillation targets for the LLM-centric RM. 
Subsequently, the LLM is trained with both the original task objective and a KL divergence loss that aligns its output distribution with that of the CRM.

Once the LLM reaches convergence, it becomes the teacher model, generating refined recommendation results for the next training phase of the CRM. The CRM now learns not only from the ground truth but also from the LLM’s recommendations via another KL divergence loss. The process alternates between freezing one model and training the other, iteratively refining both the ID embeddings and recommendation signals until the overall framework converges. This mutual distillation strategy facilitates semantic rich and robust ID embedding learning, without introducing any additional inference time or parameters.

\subsection{Downward Enhancement for LLM}
In this section, we describe how to enhance the LLM through information-augmented injection from the CRM, and introduce a sample-wise adaptive weighting mechanism to refine the knowledge distillation process.
\subsubsection{Information-Augmented Injection}
In this stage, we enhance the LLM-centric RM by incorporating informative signals from the CRM. Given the pretrained CRM, we extract the updated ID embeddings $\mathbf{E}$ and the recommendation logits $\hat{y}_c \in \mathbb{R}^{|I|}$ for each training instance. These are add into the LLM-centric RM in two ways:

\textbf{1) Injection of ID embedding.} 
Following the design of E4SRec, we project the embedded ID of the items learned by the CRM into the LLM input space, as shown in Eq.~\eqref{eq:injection}.
The resulting projected embeddings serve as part of the input sequence for the LLM, enabling it to incorporate structural information directly from the conventional model.

\textbf{2) Output Distribution Distillation.} 
We further leverage the output distributions of the CRM to transfer high-level modeling signals that are not explicitly captured by the LLM. Although LLM-centric recommendation obtained ID embedding, it cannot capture the internal parameter behaviors of the CRM other than ID embedding, such as its learned activation patterns, attention mechanisms, or non-linear transformations. 
To bridge this gap, we utilize the score $\hat{y}_c \in \mathbb{R}^{|I|}$ generated by the CRM during inference as an auxiliary learning signal for the LLM. 
These scores encode rich information beyond the raw embeddings, including sequential dependencies and user behavior patterns implicitly modeled through the CRM’s architecture. 
By aligning the output distribution of the LLM-centric RM with that of the CRM, the LLM-centric RM can indirectly absorb these modeling insights.

Specifically, we apply temperature scaling to both score vectors to produce softened probability distributions:

\begin{equation} 
\mathbf{p}_{c1} = \mathrm{Softmax}\left(\frac{\hat{y}_c}{T_1}\right), \quad 
\mathbf{p}_{l1} = \mathrm{Softmax}
\left(\frac{\hat{y}_l}{T_1}\right),
\end{equation}
where $T_1 \in (0,1)$ is the temperature parameter that plays a critical role in controlling the sharpness of these distributions. 
In recommendation tasks where the ranking of top items is critical, setting a temperature parameter $T_1 < 1$ sharpens the score distribution by amplifying differences among high-ranked items while compressing scores of lower-ranked ones. This mechanism enhances the discriminability of top items, which is essential for improving ranking performance.

Then define the KL divergence loss as:  
\begin{equation} 
\mathcal{L}_{\text{KL}}^{\downarrow} = D_{\text{KL}}\left( \mathbf{p}_{c1} \parallel \mathbf{p}_{l1} \right) 
\end{equation}  


\subsubsection{Sample-wise Adaptive Weighting Mechanism}
As the LLM-centric RM generally outperforms the CRM on most samples due to its stronger semantic modeling capabilities, we further introduce a sample-wise adaptive weighting mechanism to refine the KL divergence loss. The main idea is to dynamically adjust the weight of the KL loss based on the relative performance of the two models on each individual sample.

Specifically, for each training sample, we evaluate the rank of the target item in the prediction lists produced by both the CRM and the LLM-based recommendation model. Let $ r_c $ and $ r_l $ denote the corresponding ranks in the CRM and LLM outputs, respectively.
If the LLM-centric RM ranks the target item higher than the CRM (i.e., $ r_l < r_c $), it indicates that the LLM performs well on this particular sequence, and thus we reduce the influence of the KL loss to avoid unnecessary interference. 
Conversely, if the CRM achieves a better ranking, we increase the KL loss weight to encourage stronger learning from the CRM.

Formally, we define the sample-wise weight $ w_1 $ as:

\begin{equation}
w_1 = 
\begin{cases}
\max(0, \min(2, 1 + \frac{r_l - r_c}{\max(r_c, 1)})), & \text{if } r_l > r_c \\
\max(0, \min(2, 1 - \frac{r_c - r_l}{\max(r_l, 1)})), & \text{otherwise}
\end{cases},
\end{equation}
which ensures that $ w_1 \in [0, 2] $ by applying clipping to the calculated weights. It dynamically adjusts the KL loss weight based on the relative ranking performance of the CRM and LLM, while keeping the values within a predefined range.

Finally, the weighted KL divergence loss is incorporated into the overall objective function of the LLM-centric RM: 

\begin{equation}
\mathcal{L}_{\text{total}}^{\text{LLM}} = \mathcal{L}_{\text{rec}}^{\text{LLM}} + \gamma \cdot w_1 \cdot \mathcal{L}_{\text{KL}}^{\downarrow},
\end{equation}
where $ \gamma $ is a balancing coefficient.

This adaptive distillation strategy ensures that the LLM benefits from the CRM's strengths without being overly constrained on samples where it already performs well.

\subsection{Upward Semantics Distillation for CRM}
In this stage, we transfer the semantic modeling strengths of the LLM to the CRM through an upward distillation process.

\subsubsection{Output Distribution Distillation}
Similar to Section 4.2, we utilize the output distributions generated by the LLM as auxiliary supervision signals for the CRM. Specifically, for each training instance, we obtain the recommendation logits $\hat{y}_l \in \mathbb{R}^{|I|}$ from the LLM-centric RM and apply temperature scaling to produce softened probability distributions:
\begin{equation}
\mathbf{p}_{l2} = \mathrm{Softmax}\left(\frac{\hat{y}_l}{T_2}\right), \quad 
\mathbf{p}_{c2} = \mathrm{Softmax}\left(\frac{\hat{y}_c}{T_2}\right),
\end{equation}
where $T_2 \in (0,1)$ is a temperature parameter used to sharpen the distribution for better discriminability among top items.

The KL divergence loss is then defined as:
\begin{equation}
\mathcal{L}_{\text{KL}}^{\uparrow} = D_{\text{KL}}\left( \mathbf{p}_{l2} \parallel \mathbf{p}_{c2} \right),
\end{equation}
which encourages the CRM to align its output distribution with that of the LLM, thereby absorbing high-level semantic insights that the CRM may not capture on its own.

\subsubsection{Sample-wise Adaptive Weighting Mechanism}

To ensure that the CRM learns effectively without being misled by potentially suboptimal LLM predictions, we again adopt a sample-wise adaptive weighting mechanism. Formally, we define the weight $ w_2 $ as:

\begin{equation}
w_2 = 
\begin{cases}
\max(0, \min(2, 1 + \frac{r_c - r_l}{\max(r_l, 1)})), & \text{if } r_c > r_l \\
\max(0, \min(2, 1 - \frac{r_l - r_c}{\max(r_c, 1)})), & \text{otherwise}
\end{cases}.
\end{equation}

This ensures that $ w_2 \in [0, 2] $ and allows the CRM to learn selectively from the LLM based on their relative performance per sample.

Finally, the weighted KL loss is incorporated into the CRM’s objective function:

\begin{equation}
\mathcal{L}_{\text{total}}^{\text{CRM}} = \mathcal{L}_{\text{rec}}^{\text{CRM}} + \beta \cdot w_2 \cdot \mathcal{L}_{\text{KL}}^{\uparrow},
\end{equation}
where $ \beta $ is a balancing coefficient.

As more semantic knowledge is distilled into the CRM, it gains improved ID embeddings and output distributions. This, in turn, raises the performance ceiling of the LLM in the next iteration, forming a virtuous cycle of mutual enhancement between the two models.

\subsection{Training \& Deployment}

We adopt an iterative training framework that alternates between enhancing the LLM using signals from the CRM and refining the CRM with semantic knowledge distilled from the LLM. The overall process is summarized as follows:

\begin{algorithm}[htbp]
\caption{Mutual Distillation Training Procedure for \myString}
\begin{algorithmic}[1]
\Statex \textbf{Input:} Interaction dataset $\mathcal{D}$, Instruction dataset $\mathcal{D}_{\text{ins}}$, parameters $\gamma$, $\beta$, temperatures $T_1$, $T_2$, max iterations $T_{\max}$
\Statex \textbf{Output:} CRM $\mathcal{M}_{\text{CRM}}$ and LLM-centric RM $\mathcal{M}_{\text{LLM}}$

\State Pretrain CRM $\mathcal{M}_{\text{CRM}}$ on $\mathcal{D}$ to obtain item embeddings $\mathbf{E}$

\State Initialize LLM projection layers $W_{\text{in}}, W_{\text{out}}$ and LoRA modules

\For{$t \gets 0$ to $T_{\max}$ iterations}

    {\textit{// Step 1}}
    \State 
    Project the ID embeddings of CRM to obtain the ID embeddings for LLM:
    \[
    \mathbf{F}_{\text{proj}} = W_{\text{in}} \cdot \mathbf{E}
    \]
    \State Integrate $\mathbf{F}_{\text{proj}}$ into LLM token layer.
    
    \textit{// Step 2}
    \State Freeze CRM
    \State Fine-tune LLM-centric RM until convergence by loss:
    \[
    \mathcal{L}_{\text{LLM}} = \mathcal{L}_{\text{rec}}^{\text{LLM}} + \gamma \cdot w_1 \cdot \mathcal{L}_{\text{KL}}^{\downarrow}(T_1)
    \]
    
    \textit{// Step 3}
    \State Freeze LLM-centric RM
    \State Train CRM until convergence by loss:
    \[
    \mathcal{L}_{\text{CRM}} = \mathcal{L}_{\text{rec}}^{\text{CRM}} + \beta \cdot w_2 \cdot \mathcal{L}_{\text{KL}}^{\uparrow}(T_2)
    \]
\EndFor
\Statex \textit{// Model Deployment}
\end{algorithmic}
\end{algorithm}

Owing to the design of our mutual distillation framework, which focuses on information flow and loss-level interactions without introducing additional trainable components into either the CRM or the LLM backbone, our method brings negligible overhead to the inference stage. 
During deployment, the CRM and the LLM-centric RM can be independently deployed, depending on the specific application scenario. Since no structural changes are made to the original models, they can be integrated into existing serving infrastructures with minimal modification. This design ensures that the improved performance does not come at the cost of increased inference latency or resource consumption.

\FloatBarrier

\section{EXPERIMENTS} 
In this section, we conduct comprehensive experiments to evaluate the effectiveness of our proposed mutual distillation framework \myString, which enables bidirectional knowledge transfer between CRMs and a LLM-centric RMs. We benchmark our framework on several real-world datasets and compare it with representative baselines in sequential recommendation. 
Moreover, we conduct ablation studies, efficiency evaluations, etc. to answer the following research questions:  
\begin{itemize}     
\item 
    \textbf{RQ1:} 
    How does our mutual distillation framework perform compared with standard CRMs and LLM-centric RMs?
    \item \textbf{RQ2:} 
    What are the individual contributions of different components?
    \item \textbf{RQ3:} 
    How do the performances of the CRM and LLM-centric RM evolve as training progresses over multiple rounds?
    \item \textbf{RQ4:} 
    How do training and inference latencies of the proposed framework compare in practical deployment settings?
    
\end{itemize}

\subsection{Settings}
\subsubsection{Datasets}
To assess the performance of our proposed framework, we utilize four commonly adopted real-world datasets. 
Specifically, we utilize three Amazon categories~\cite{mcauley2015image}: "Beauty", "Sports and Outdoors", "Toys and Games", as well as the Yelp dataset~\cite{ni2019justifying}, which contains user interaction histories with local businesses.
To ensure the recency and relevance of the data, we apply a temporal filter to all datasets and retain only interactions that occurred on or after January 1st, 2019.
For sequential recommendation tasks, we sort each user's interactions chronologically. 
Consistent with prior studies ~\cite{sun2019bert4rec, li2023e4srec}, we adopt the 5-core setting, removing items with fewer than five interactions to reduce data sparsity and enhance evaluation reliability. Dataset statistics are summarized in Table~\ref{tab:dataset-stats}.

\subsubsection{Baselines.}
We compare our proposed method with two categories of sequential recommendation approaches:
(1) \textbf{CRM-centric methods}, where conventional sequential recommendation models serve as the core recommenders. This category includes both original model-based methods and those enhanced by LLMs such as knowledge distillation or prompt-guided augmentation.
(2)\textbf{LLM-centric methods}, where LLMs themselves take the role of the main recommendation module, generating recommendation results in an end-to-end fashion.

CRM-centric methods include:

\begin{itemize}     
\item 
    \textbf{GRU4Rec}~\cite{hidasi2015session} leverages a gated recurrent unit (GRU) network to learn user behavior sequences and make recommendation predictions.
    \item \textbf{SASRec}~\cite{kang2018self} utilizes a self-attention mechanism with multi-head design to capture sequential dependencies and generate recommendations accordingly.
    \item \textbf{S$^3$-Rec}~\cite{zhou2020s3} applies contrastive learning to uncover correlations among items, sub-sequences, and attributes. We use the version with only the Masked Item Prediction (MIP) objective.
     \item \textbf{DLLM2Rec}~\cite{cui2024distillation} is a framework that transfers knowledge from LLM-centric RMs to CRMs by addressing teacher reliability, capacity gap, and semantic mismatch via importance-aware ranking and collaborative embedding distillation.
    \item \textbf{LLM-CF}~\cite{sun2024large} is a framework that distills the knowledge and reasoning capabilities of LLMs into collaborative filtering via in-context learning and instruction tuning.
    \item \textbf{LEADER}~\cite{liu2024large} introduces a feature-level knowledge distillation approach that transfers the semantic understanding of LLMs to lightweight models, improving efficiency and accuracy in medication recommendation. In our work, we focus solely on its distillation component.
\end{itemize}

\begin{table}[t]
\small
  \centering
  \caption{Statistics of datasets}
  \begin{tabular}{lrrrrr}
    \toprule
    Dataset & \# Users & \# Items & \# Interactions & Density \\
    \midrule
    Beauty & 22,363 & 12,101 & 198,502 & 0.07\% \\
    Sports & 25,598 & 18,357 & 296,337 & 0.06\% \\
    Toys & 19,412 & 11,924 & 167,597 & 0.07\% \\
    Yelp & 30,431 & 20,033 & 316,354 & 0.05\% \\
    \bottomrule
  \end{tabular}

    \label{tab:dataset-stats}
\end{table}

\begin{table*}[htbp]
\centering
\caption{Performance comparison of different methods. For fair evaluation, we report the results of CRM-centric methods and LLM-centric methods separately. The best result for each metric is bolded, and the second best is underlined. 'Imp.s' indicates the relative improvement of our framework on the corresponding backbone (SASRec or E4SRec), while 'Imp.b' indicates the improvement over the best baseline within the same category.}
\renewcommand{\arraystretch}{1.2}
\begin{adjustbox}{max width=\textwidth}
\begin{tabular}{l l|cccccc|ccc|cc|ccc}
\toprule
\multirow{2}{*}{\textbf{Dataset}} & \multirow{2}{*}{\textbf{Metric}} & \multicolumn{9}{c|}{\textbf{CRM-centric methods}} & \multicolumn{5}{c}{\textbf{LLM-centric methods}} \\

\cmidrule(r){3-11} \cmidrule(r){12-16}
& & GRU4Rec  & SASRec & S$^3$Rec & DLLM2Rec & LLM-CF & LEADER & \myStringC  & Imp.s & Imp.b & E4SRec & LLM-SRec & \myStringL & Imp.s & Imp.b \\
\midrule

\multirow{6}{*}{\textbf{Beauty}} 
& HR@5  &0.0165 & 0.0260 & 0.0262 & 0.0316 & \underline{0.0377} & 0.0299 & \textbf{0.0453} & 74.23\% & 20.16\% & \underline{0.0518} & 0.0510 & \textbf{0.0553} & 6.76\% & 6.76\% \\

& HR@10  & 0.0286 & 0.0504 & 0.0483 & 0.0553 & \underline{0.0632} & 0.0516 & \textbf{0.0758} & 50.40\% & 19.94\% & 0.0749 & \underline{0.0756} & \textbf{0.0814} & 8.68\% & 7.67\% \\

& HR@20  & 0.0526 & 0.0855 & 0.0816 & 0.0884 & \underline{0.1013} & 0.0870 & \textbf{0.1157} & 35.32\% & 14.22\% & 0.1069 & \underline{0.1092} & \textbf{0.1150} & 7.58\% & 5.31\% \\

& NDCG@5 & 0.0102 & 0.0125 & 0.0133 & 0.0167 & \underline{0.0216} & 0.0158 & \textbf{0.0256} & 104.80\% & 18.52\% & \underline{0.0360} & 0.0345 & \textbf{0.0379} & 5.28\% & 5.28\% \\

& NDCG@10 & 0.0141 & 0.0204 & 0.0206 & 0.0243 & \underline{0.0298} & 0.0228 & \textbf{0.0354} & 73.53\% & 18.79\% & \underline{0.0434} & 0.0424 & \textbf{0.0463} & 6.68\% & 6.68\% \\

& NDCG@20 & 0.0201 & 0.0292 & 0.0290 & 0.0326 & \underline{0.0384} & 0.0317 & \textbf{0.0454} & 55.48\% & 18.23\% & \underline{0.0515} & 0.0508 & \textbf{0.0547} & 6.21\% & 6.21\% \\

\midrule 
\multirow{6}{*}{\textbf{Sports}} 

& HR@5  & 0.0110 & 0.0173 & 0.0158 & 0.0181 & \underline{0.0216} & 0.0174 & \textbf{0.0284} & 52.16\% & 33.96\% & 0.0245 &  \underline{0.0286} & \textbf{0.0304} & 24.08\% & 6.29\% \\

& HR@10  &0.0187 & 0.0307 & 0.0261 & 0.0333 & \underline{0.0382} & 0.0313 & \textbf{0.0468} & 62.44\% & 22.51\% & 0.0373 &  \underline{0.0474} & \textbf{0.0474} & 27.08\% & 9.98\% \\

& HR@20  & 0.0312 & 0.0497 & 0.0383 & 0.0521 & \underline{0.0603} & 0.0518 & \textbf{0.0715} & 43.86\% & 17.41\% & 0.0557 & \underline{0.0660} & \textbf{0.0692} & 24.24\% & 4.85\% \\

& NDCG@5 & 0.0068 & 0.0091 & 0.0094 & 0.0097 & \underline{0.0120} & 0.0096 & \textbf{0.0162} & 78.02\% & 33.88\% & 0.0168 & \underline{0.0184} & \textbf{0.0197} & 17.26\% & 7.07\% \\

& NDCG@10 & 0.0092 & 0.0134 & 0.0121 & 0.0144 & \underline{0.0176} & 0.0135 & \textbf{0.0221} & 64.93\% & 25.57\% & 0.0211 & \underline{0.0235} & \textbf{0.0252} & 19.43\% & 7.23\% \\

& NDCG@20 & 0.0123 & 0.0182 & 0.0164 & 0.0191 & \underline{0.0230} & 0.0187 & \textbf{0.0283} & 55.49\% & 26.91\% & 0.0256 & \underline{0.0271} & \textbf{0.0307} & 19.92\% & 13.28\% \\

\midrule 
\multirow{6}{*}{\textbf{Toys}} 
& HR@5  & 0.0158 & 0.0427 & 0.0378 & 0.0447 & \underline{0.0465} & 0.0457 & \textbf{0.0552} & 29.27\% & 18.71\% & \underline{0.0553} & 0.0552 & \textbf{0.0606} & 9.58\% & 9.58\% \\

& HR@10  & 0.0275 & 0.0664 & 0.0615 & 0.0708 & 0.0757 & \underline{0.077} & \textbf{0.0815} & 22.74\% & 5.84\% & 0.0776 & \underline{0.0814} & \textbf{0.0889} & 14.56\% & 9.21\% \\

& HR@20  & 0.0454 & 0.0989 & 0.0914 & 0.1047 & 0.1084 & \underline{0.1143} & \textbf{0.1189} & 20.22\% & 4.02\% & 0.1072 &  \underline{0.1145} & \textbf{0.1211} & 12.97\% & 5.76\% \\

& NDCG@5 &  0.0095 & 0.0221 & 0.0199 & 0.0231 & 0.0251 & \underline{0.0273} & \textbf{0.0340} & 53.85\% & 24.54\% & \underline{0.0403} & 0.0385 & \textbf{0.0425} & 5.46\% & 5.46\% \\

& NDCG@10 & 0.0132 & 0.0297 & 0.0275 & 0.0316 & 0.0341 & \underline{0.0354} & \textbf{0.0398} & 34.01\% & 12.43\% & 0.0475 & \underline{0.0479} & \textbf{0.0516} & 8.63\% & 7.72\% \\

& NDCG@20 & 0.0177 & 0.0378 & 0.0352 & 0.0401 & 0.0423 & \underline{0.0451} & \textbf{0.0495} & 30.95\% & 9.76\% & 0.0549 & \underline{0.0562} & \textbf{0.0598} & 8.93\% & 6.41\% \\

\midrule 
\multirow{6}{*}{\textbf{Yelp}} 
& HR@5  & 0.0196 & 0.0223 & 0.0237 & 0.0259 & \underline{0.0275} & 0.0252 & \textbf{0.0391} & 75.34\% & 42.18\% & 0.0309 & \underline{0.0316} & \textbf{0.0366} & 18.45\% & 15.82\% \\

& HR@10  & 0.0326 & 0.0390 & 0.0379 & 0.0404 & 0.0417 & \underline{0.0451} & \textbf{0.0626} & 60.51\% & 38.80\% & 0.0478 &  \underline{0.0526} & \textbf{0.0582} & 21.76\% & 10.68\% \\

& HR@20  & 0.0543 & 0.0661 & 0.0604 & 0.0662 & 0.0654 & \underline{0.0730} & \textbf{0.0985} & 49.02\% & 34.93\% & 0.0735 & \underline{0.0851} & \textbf{0.0901} & 22.59\% & 5.88\% \\

& NDCG@5 & 0.0121 & 0.0141 & 0.0154 & 0.0179 & \underline{0.0190} & 0.0159 & \textbf{0.0257} & 82.27\% & 35.26\% & \underline{0.0211} & 0.0199 & \textbf{0.0244} & 15.64\% & 15.64\% \\

& NDCG@10 & 0.0162 & 0.0194 & 0.0201 & 0.0225 & \underline{0.0236} & 0.0223 & \textbf{0.0332} & 71.13\% & 40.68\% & 0.0265 & \underline{0.0274} & \textbf{0.0314} & 18.49\% & 14.60\% \\

& NDCG@20 & 0.0217 & 0.0262 & 0.0258 & 0.0287 & 0.0291 & \underline{0.0294} & \textbf{0.0423} & 61.45\% & 43.88\% & 0.0330 & \underline{0.0352} & \textbf{0.0393} & 19.09\% & 11.65\% \\

\bottomrule
\end{tabular}
\end{adjustbox}
\label{tab:exp1}
\end{table*}

LLM-centric methods include:

\begin{itemize}
    \item \textbf{E4SRec}~\cite{li2023e4srec} is a method that combines LLMs with ID-based recommendation by using item ID sequences as input and generating rankings efficiently with minimal trainable parameters, improving practicality and scalability.
    
    \item \textbf{LLM-SRec}~\cite{kim2025lost} is a method that distills user representations from a pretrained sequential model into LLMs to better integrate sequential information for recommendation.
\end{itemize}

Some existing LLM-centric recommendation models rely on sampling a limited number of negative samples per positive instance to form candidate sets for scoring, due to their inability to generate scores for all items simultaneously. In contrast, our model generates scores for all items in a single pass. To ensure a fair comparison, we do not include these sampling-based methods in our evaluation.

\subsubsection{Evaluation Metrics.}
Following common practices~\cite{xu2024slmrec,li2023e4srec,zhou2020s3,zhou2022filter}, we use the leave-one-out strategy to partition the training, validation, and test datasets. We consider all recorded interactions in the datasets as positive samples and randomly sample a number of negative items for each user. For evaluation, we adopt widely used Top@K metrics, including Hit Ratio (HR@K) and Normalized Discounted Cumulative Gain (NDCG@K), which measure whether the ground-truth item is ranked among the top-K recommended items and how high it is ranked, respectively.

\subsubsection{Implementation Details.}
For GRU4Rec, SASRec, S$^3$-Rec, LLM-CF, and E4SRec, we implement the models using the public resources provided by the authors. For other baselines, we build their distillation components based on SASRec and E4SRec using PyTorch 2.2.1. In our distillation framework, we adopt SASRec and E4SRec as the CRM-based and LLM-centric base recommenders, respectively, due to their widespread use in prior work.
Specifically, for E4SRec, we adopt Qwen2.5-7B as the underlying LLM, obtained from HuggingFace\footnote{\url{https://huggingface.co/Qwen/Qwen2.5-7B-Instruct}}, and initialize the ID embeddings using pretrained SASRec. 
We fine-tune E4SRec with a LoRA module~\cite{hu2022lora} of rank 8. The total batch size under gradient accumulation is set to 128, and the learning rate is fixed at $3 \times 10^{-4}$.
Across all datasets and models, we set the embedding dimension to 64 and the maximum sequence length to 50. The Adam optimizer~\cite{kingma2014adam} is used for optimization. 
In our proposed framework, we set the temperature parameters $T_1$ and $T_2$ to 0.6 and 0.2, respectively, and the maximum number of iterative refinement steps $T_{\text{max}}$ to 2. 
We also set the balancing coefficients $ \gamma $  and $ \beta $ to 1 and 0.5. All models are trained and evaluated on two Nvidia A800 GPUs.

\subsection{Performance Comparison (RQ1)} 
The performance comparison of our \myString{} with other sequential recommendation models is shown in Table~\ref{tab:exp1}. Here we have the following observations:

\textbf{Overall:} Our proposed mutual distillation framework, including \myStringC{} (SASRec backbone) and \myStringL{} (E4SRec backbone), significantly outperforms all baselines on four datasets, thanks to effective collaboration between CRM and LLM-centric RM.
\myStringC{} achieves an average improvement of approximately 22.7\% on HR@k and 25.7\% on NDCG@k over the best baselines in its category, while \myStringL{} shows gains of around 8.2\% on HR@k and 8.9\% on NDCG@k. 
On three Amazon datasets (Beauty, Sports and Toys), \myStringL{} generally performs better than \myStringC{}, especially in ranking metrics such as NDCG@k. On Yelp, however, \myStringC{} outperforms \myStringL{}, indicating stronger learning capability in that setting, potentially due to simpler user behavior patterns or reduced noise in LLM integration.

\textbf{\myStringC{} compared with CRMs (GRU4REC, SASRec, S$^3$Rec):} 
    \myStringC{} consistently outperforms these CRMs across all datasets' evaluation metrics. This indicates that distilling knowledge from LLMs significantly enhances the model’s overall performance. The distilled information allows \myStringC{} to better understand user preferences and provide more accurate personalized scores. By integrating knowledge from LLMs, \myStringC{} can more effectively capture nuanced patterns in user behavior, leading to improved recommendation accuracy.

\textbf{\myStringC{} compared with LLM-enhanced CRMs 
 (LLM-CF, DLLM2Rec, LEADER):} 
    Despite the advantage of LLM-CF of incorporating textual features of items and users, and LEADER and DLLM2Rec's strategy of distilling partial components from LLM-centric models, \myStringC{} still demonstrates competitive and superior performance. This highlights the effectiveness of our framework in capturing more informative and well-structured knowledge through distribution modeling and mutual distillation loops.

\textbf{\myStringL{} compared with LLM-centric methods (E4SRec, LLM-SRec):} 
    \myStringL{} consistently outperforms E4SRec and LLM-SRec because the mutual distillation framework allows the model to learn structural knowledge from CRM during the iterative distillation process. Additionally, \myStringL{} benefits from more refined ID embeddings obtained through optimized CRMs, leading to enhanced generalization and recommendation performance.

\begin{figure*}[htbp]       
\centering       
\includegraphics[width=1.0\textwidth]{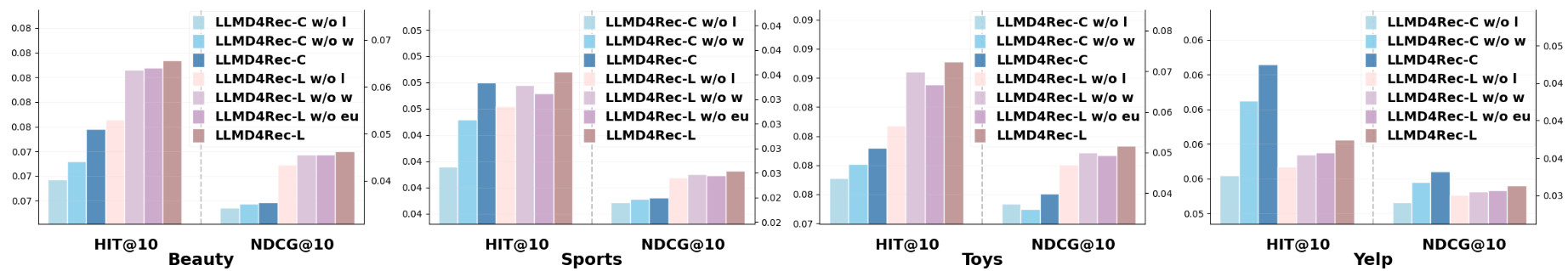}
\caption{Ablation study of the loop mechanism, distillation sample weight, and ID embedding update in \myString{}.
}
\label{fig:exp2} 
\end{figure*}

\begin{figure}[t]        
    \includegraphics[width=0.99\columnwidth]{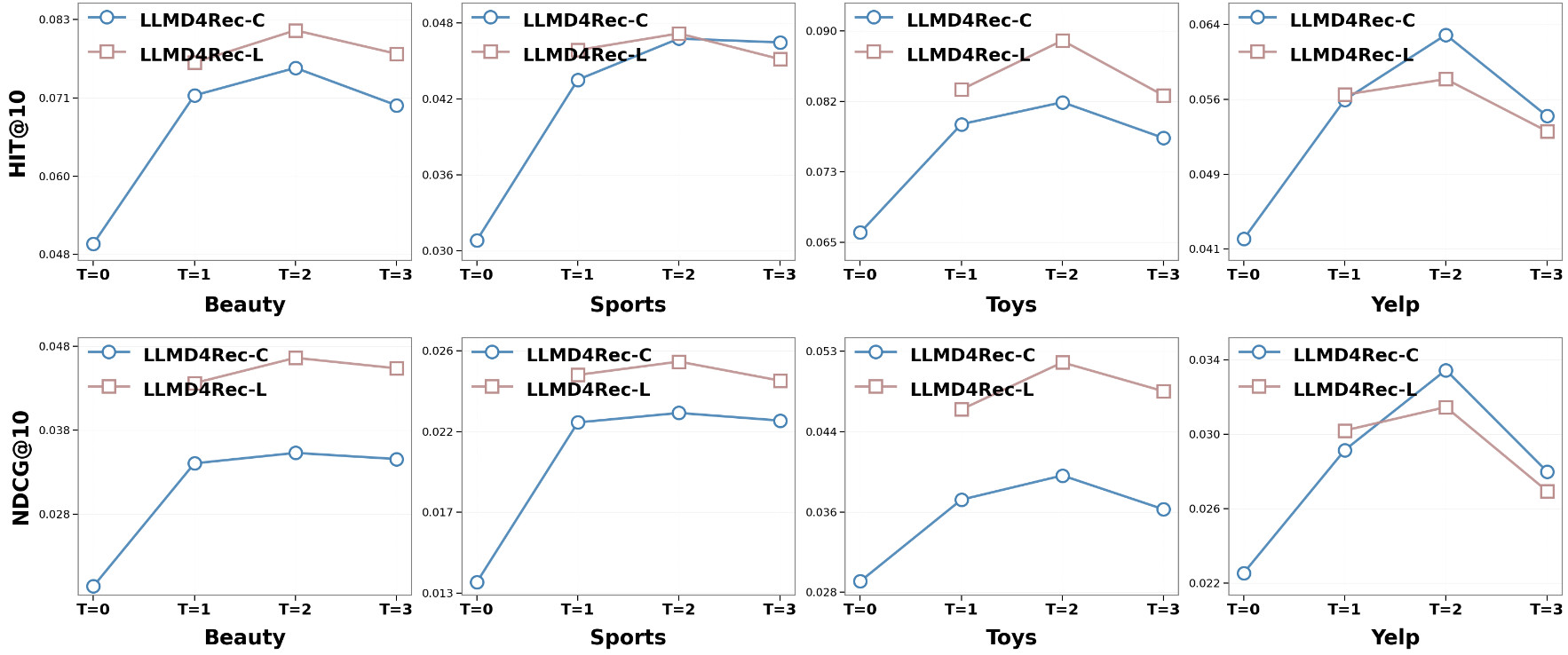}     
    \caption{Overview of the proposed \myString{} framework
    }       
\label{fig:exp3} 
\end{figure}
\subsection{Ablation Study (RQ2)}
In ablation study, we perform a thorough analysis of our framework to evaluate the contribution of each component. 
Specifically, we tested the CRM-centric model \myStringC{} and the LLM-centric model \myStringL{} in different configurations.

For \myStringC{}, we examine two variants: \myStringC{} w/o l (loop) and \myStringC{} w/o w (sample weight).
As shown in Figure~\ref{fig:exp2}, \myStringC{} significantly outperforms the other two variants. 
This demonstrates the critical role of the loop mechanism and distillation sample importance weights in enhancing recommendation result. 
Notably, the performance of \myStringC{} w/o l is considerably lower than that of \myStringC{}, indicating that the loop mechanism helps optimize model parameters gradually. 
Similarly, \myStringC{} w/o w performs worse than \myStringC{}, demonstrating that sample-wise adaptive weighting mechanism can effectively guide model training and further enhance recommendation quality.

Next, we analyze the LLM-centric model \myStringL{}, including its variants: \myStringL{} w/o l (loop), \myStringL{} w/o w (sample weight) and \myStringL{} w/o eu (embedding update). 
Experimental results show that \myStringL{} achieves the best performance across all metrics, confirming its effectiveness in complex scenarios. The removal of the loop or weight components leads to performance drops, highlighting their importance. Additionally, disabling embedding updates slightly degrades performance, indicating that updating ID embeddings at the beginning of each loop is valuable for model adaptability.

In conclusion, the ablation study confirms that the loop mechanism, distillation sample importance weights, and dynamic ID embedding updates are beneficial to the performance of our \myString{} method.

\subsection{Performance over Training Rounds (RQ3)}
Figure~\ref{fig:exp3} presents the performance trends of HIT@10 and NDCG@10 for both models as the number of training rounds increases. At $T = 0$, \myStringC{} is initialized and trained as a standalone SASRec model. Starting from $T \geq 1$, both \myStringC{} and \myStringL{} engage in an iterative mutual distillation process, resulting in significant performance improvements across all four datasets. Notably, after the first round of training ($T = 1$), both components exhibit substantial gains, indicating that mutual distillation effectively enhances their learning efficiency. At $T = 2$, \myStringC{} and \myStringL{} achieve optimal performance on most metrics, demonstrating that the proposed framework achieves incremental improvements through continuous knowledge exchange.

\subsection{Efficiency Analysis (RQ4)}
Given that our approach does not introduce any additional trainable parameters but merely incorporates two auxiliary losses, there are no significant impacts on either inference speed or storage requirements. This characteristic makes our method highly efficient without compromising model performance.
The following table provides a comparative overview of E4SRec and \myStringL{} regarding training efficiency.

\begin{table}[t]\small
\caption{Comparison of Training Efficiency between E4SRec and One-Loop \myStringL}
\centering
\begin{tabular}{lccc}
\toprule
\textbf{Method} & \textbf{Dataset} &\textbf{Training Time (s)} & \textbf{Hit@10}\\
\midrule
\midrule
\multirow{2}{*}{E4SRec} & Beauty & 1.3×10$^4$ & 0.0749 \\
                       & Toys    & 1.1×10$^4$ & 0.0766 \\
\cmidrule(lr){1-4}
\multirow{2}{*}{\myStringC{}} & Beauty & 9.0×10$^3$ & 0.0776 \\
                              & Toys   & 7.4×10$^3$ & 0.0831 \\
\bottomrule
\end{tabular}
\label{tab:comparison}
\end{table}

Notably, for E4SRec, which typically requires a longer training duration, we expedite its training process by integrating traditional models' scoring mechanisms. This adjustment enables us to achieve superior results within just two epochs during the first cycle, whereas the backbone averages three epochs to reach comparable performance. As shown in Table~\ref{tab:comparison}, this improvement significantly reduces the overall training time while maintaining high levels of accuracy.

\section{Conclusions}
We propose \myString, a mutual distillation framework that enables bidirectional and dynamic knowledge transfer between CRMs and LLM-centric RMs. By aligning output distributions and introducing a sample-wise adaptive weighting mechanism, the framework maximizes knowledge utilization without additional parameters, effectively combining the strengths of CRMs in modeling collaborative signals and LLMs in capturing deep semantic patterns. 
We hope that this work will significantly advance the state-of-the-art in recommendation systems by providing a robust and versatile solution for integrating diverse models.
Future work may explore multi-modal integration, adaptive strategies for complex scenarios, reinforcement learning-optimized distillation loops, efficient parameter sharing, and real-time recommendation applications to further expand its utility.

\section{Generative AI Disclosure}

Generative Al tools were used solely for language refinement after the main content of the paper was completed. No GenAI tools were used in the design, implementation, analysis, or writing of the research content itself.

\bibliographystyle{ACM-Reference-Format}
\bibliography{sample-base}

\appendix

\end{document}